\documentclass[amsmath,amssymb,aps,twocolumn,10pt,journal=pra,longbibliography]{revtex4-2}
\usepackage{amssymb}
\usepackage{graphicx}
\usepackage{dcolumn}
\usepackage{bm}
\usepackage{amsmath}
\usepackage{subfigure}
\usepackage{color}
\usepackage{graphicx}
\usepackage[colorlinks=true,citecolor=blue]{hyperref}

\begin{document}

%\preprint{APS/123-QED}

\title{\texorpdfstring{Measurements of the hyperfine structure of $n\mathrm{P_J}$ Rydberg states by microwave spectroscopy in Cs atoms}{Measurements of the hyperfine structure of nP(J) Rydberg states by microwave spectroscopy in Cs atoms}}

\author{Rong Song$^{1}$}
\author{Jingxu Bai$^{1,2}$}
\author{Zhenhua Li$^{1}$}
\author{Yuechun Jiao$^{1,2}$}
\thanks{ycjiao@sxu.edu.cn}
\author{Suotang Jia$^{1,2}$}
\author{Jianming Zhao$^{1,2}$}
\thanks{zhaojm@sxu.edu.cn}

\affiliation{
$^{1}$State Key Laboratory of Quantum Optics and Quantum Optics Devices, Institute of Laser Spectroscopy, Shanxi University, Taiyuan 030006, People's Republic of China\\
$^{2}$Collaborative Innovation Center of Extreme Optics, Shanxi University, Taiyuan 030006, People's Republic of China}

\date{\today}% It is always \today, today,

\begin{abstract}
 We present measurements of hyperfine structure (HFS) of the $n\mathrm{P_J}$ Rydberg states for large principal quantum number $n$ range ($n=41-55$)  employing the microwave spectroscopy in an ultra-cold cesium Rydberg ensemble. A microwave field with 30-$\mu$s duration couples the $ n\mathrm{S} \to n\mathrm{P} $ transition, yielding a narrow linewidth spectroscopy that approaches the Fourier limit,  which allows us to resolve the hyperfine structure of $ n\mathrm{P_J} $ states. By analyzing the hyperfine splittings of $n\mathrm{P_{J}}$ states, 
 we determine the magnetic-dipole HFS coupling constant $\Bar{\mathrm{A}}_{\mathrm{HFS,P_{1/2}}}=3.760(26) $~GHz for $\mathrm{P_{1/2}}$ state,   $\Bar{\mathrm{A}}_{\mathrm{HFS,P_{3/2}}}= 0.718(27)$~GHz, and $ \Bar{\mathrm{B}}_{\mathrm{HFS,P_{3/2}}}= -0.084(102)$~GHz for $\mathrm{P_{3/2}}$ state, respectively. 
Systematic uncertainties caused by stray electromagnetic field, microwave field power and Rydberg interaction are analyzed. 
This measurement is significant for the investigation of Rydberg electrometry and quantum simulation with dipole interaction involving $n\mathrm{P_J}$ state.

\end{abstract}

%\keywords{}

\maketitle

%\tableofcontents

\section{Introduction}
\label{subsec:intro}

Precise measurements of the fine and hyperfine energy levels of Rydberg states and their quantum defects play a role in testing
atomic-structure and quantum-defect theories~\cite{seaton1983quantum}, as well as in applications of Rydberg-atom-based metrology~\cite{sedlacek2012microwavea,jiao2017atombased,anderson2021selfcalibrated} and Rydberg molecules~\cite{booth2015production,heiner2015experimental,niederprum2016observation,bai2024microwavea}.
The leading term in the hyperfine structure (HFS) is the magnetic-dipole interaction between nuclear and electronic angular momenta, followed by the interaction between nuclear electric quadrupole moment and electronic electric-field gradient.
Hyperfine splittings of $n\mathrm{P_J}$ Rydberg states are dozens of kHz, such as the hyperfine splitting of 41$\mathrm{P_{3/2} (F'=3) }$and $\mathrm{(F'=4)}$ 
is only $\sim$50~kHz, which cannot be resolved with most laser-spectroscopic methods. 
In contrast, microwave spectroscopy of transitions between Rydberg states routinely allows resolutions in the range of tens of kHz, often limited only by the atomic lifetime and the atom-field interaction time.
Here, we employ high-resolution microwave spectroscopy to measure the HFS of high-lying Cs $n\mathrm{P_J}$ Rydberg states.

Measurements of the HFS of alkali atoms,
recently reviewed by M. Allegrini {\sl{et al.}}~\cite{allegrini2022survey}, have been performed over a wide range of the principal quantum number $n$ and for 
several angular momenta $l$.
Laser spectroscopy in Cs vapor cells was employed to measure the Cs HFS 
for $n \lesssim 18$ and $l = 0-3 $ (see~\cite{allegrini2022survey%,deech1977lifetimes
} and references therein). 
A measurement of the magnetic-dipole HFS coupling constant, $\mathrm{A_{HFS}}$, of Cs $6\mathrm{P_{1/2}}$ was first reported 
in~\cite{abele1975untersuchung}, and later in~\cite{coc1987isotope,abele1975bestimmung,rafac1997measurement,udem1999absolute,das2006rubidiumstabilizeda,das2006precisea,gerginov2006optical,truong2015accurate} based on different measurement methods. 
The electric-quadrupole HFS coupling constant, $\mathrm{B_{HFS}}$, of
$ 6\mathrm{P}_{3/2}$ was first measured in~\cite{thibault1981hyperfine}, with subsequent improvements reported in~\cite{tanner1988precision,gerginov2003observation,das2005hyperfine}. For $n\mathrm{P_{3/2}} (n=7-10) $ states, both $\mathrm{A_{HFS}}$ and $\mathrm{B_{HFS}}$ were 
measured~\cite{arimondo1977experimental,williams2018spectroscopica,bucka1962hyperfeinstruktur,faist1964frequency,bayram2014quantum,rydberg1972investigation}; however, for $ n\mathrm{P_{3/2}}$ with  $n \geq  10$ only the magnetic-dipole constant $\mathrm{A_{HFS}}$ has been reported.
Using high-resolution double-resonance and millimeter-wave spectroscopy, P. Goy {\sl{et al.}}~\cite{goy1982millimeterwave} investigated Rydberg excitation spectra in an atomic beam and obtained quantum defects of S, P, D and F Cs Rydberg states in the $n=23-45$ range. They also obtained fine-structure as well as HFS  coupling constants, $\mathrm{A_{HFS}}$, for $n\mathrm{S}_{1/2}$ and $n\mathrm{P}_{1/2}$ in the $n=23-28$ range. Recently, H. Saßmannshausen {\sl{et al.}}~\cite{sassmannshausen2013highresolutionb}  measured high-resolution, Fourier-limited microwave spectra of Rydberg states in an ultracold Cs gas 
and obtained HFS coupling constants $\mathrm{A_{HFS}}$ for $\mathrm{S_{1/2}}$ states with $n$ up to 90, for $\mathrm{P_{3/2}}$ states with $n$ up to 72, and for the $66\mathrm{D_{3/2,5/2}}$ states.

In the present work, we measure the HFS of Cs $n\mathrm{P_J}$ states for $n=41-55$ using the high-precision microwave spectroscopy, with the microwave field driving $n\mathrm{S}_{1/2} \to n\mathrm{P_J}$ transitions. After careful compensation of stray electric and magnetic fields,
we obtain microwave spectra with 35~kHz linewidth, which approaches the Fourier limit. We extract the HFS splittings of $n\mathrm{P_J}$ states for $\mathrm{J=1/2}$ and $3/2$, from which we determine the HFS coupling constants $\mathrm{A_{HFS}} $ and $ \mathrm{B_{HFS}}$. We provide a systematic uncertainty analysis. 
Our measurements show good agreement with the previous data~\cite{allegrini2022survey}.

\section{Methods}
\subsection{Theory }

The hyperfine structure is due to the electromagnetic multipole interaction between the nucleus and electron, which is defined as couplings between the total angular momentum and the nuclear spin. Cesium atom has one valence electron with the total angular momentum J, that is coupling of the spin S and orbital angular momentum L. The nuclear spin quantum number I = 7/2. Therefore, the hyperfine shift of a $n\mathrm{P_J}$ with hyperfine quantum number F is expressed as~\cite{danieladam2019cesium}
\begin{widetext}
	\begin{equation}
      \label{eq1}
		\begin{split}
			\Delta \mathrm{E_{HFS}} &= \frac{1}{2}\mathrm{A_{HFS}K}+\mathrm{B_{HFS}}\frac{\frac{3}{2}\mathrm{K(K+1)-2I(I+1)J(J+1)}}{4\mathrm{I(2I-1)J(2J-1)}}\\
			& +\mathrm{C_{HFS}}\frac{5\mathrm{K^2(K/4+1)+K[I(I+1)+J(J+1)+3-3I(I+1)J(J+1)]-5I(I+1)J(J+1)}}{\mathrm{I(I-1)(2I-1)J(J-1)(2J-1)}}\\
		\end{split}
	\end{equation}
\end{widetext}
and \\
\begin{equation}
\label{eq2}
	\mathrm{K=F(F+1)-I(I+1)-J(J+1)}, 
\end{equation}
where $\mathrm{A_{HFS}}$ is the magnetic-dipole constant describing the magnetic dipole-dipole interaction
between the nucleus and Rydberg electron, $\mathrm{B_{HFS}}$ is the electric-quadrupole constant describing the nuclear electric-quadrupole interaction, and $\mathrm{C_{HFS}}$ is the magnetic-octupole constant representing the 
magnetic-octupole interactions between the two particles. In general, $\mathrm{C_{HFS}}$ is too small to measure in this type of experiment. For $n\mathrm{P}_{1/2}$ state, only $\mathrm{A_{HFS}}$ is nonzero. To avoid confusion, we add the subscript to distinguish the hyperfine splitting and hyperfine coupling constant $\mathrm{A_{HFS}}$ for $\mathrm{P_{1/2}}$ and $\mathrm{P_{3/2}}$ state.  In addition, for easily comparing, we introduce the reduced hyperfine coupling constant $\mathrm{A_{HFS, P_J}}$ and $\mathrm{B_{HFS, P_J}}$, and use these notations from now on.

Considering the short-range interactions scaling as $(n-\delta(n))^{-3}$~\cite{Gallagher1994}, for $n\mathrm{P_{1/2}}$ state, the hyperfine splitting $ \nu_{34,\mathrm{P}_{1/2}}$ between $\mathrm{F}'$ = 3 and 4 can be expressed as
\begin{align}
\nu_{34,\mathrm{P}_{1/2}} =\frac{4\mathrm{A_{HFS,P_{1/2}}}}{(n-\delta_n)^3}.
\label{eq:A}
\end{align}

For $n\mathrm{P_{3/2}}$ state, the hyperfine splitting between $\mathrm{F}'$=3 and 4, $\mathrm{F}'$=4 and 5 can be respectively expressed as 
\begin{align}
	\nu_{34,\mathrm{P}_{3/2}} &=\frac{4\mathrm{A_{HFS,P_{3/2}}}}{(n-\delta_n)^3}-\frac{2\mathrm{B_{HFS,P_{3/2}}}}{7(n-\delta_n)^3} \\
	\nu_{45,\mathrm{P}_{3/2}} &=\frac{5\mathrm{A_{HFS,P_{3/2}}}}{(n-\delta_n)^3}+\frac{5\mathrm{B_{HFS,P_{3/2}}}}{7(n-\delta_n)^3}. 
	\label{eq:ryd3}
\end{align}
  
From measured microwave spectroscopy, we can extract the hyperfine splitting and further the hyperfine coupling constant.

\subsection{Experimental Setup}
\label{subsec:setup}

\begin{figure}[htbp]
	\centering\includegraphics[width=0.5\textwidth]{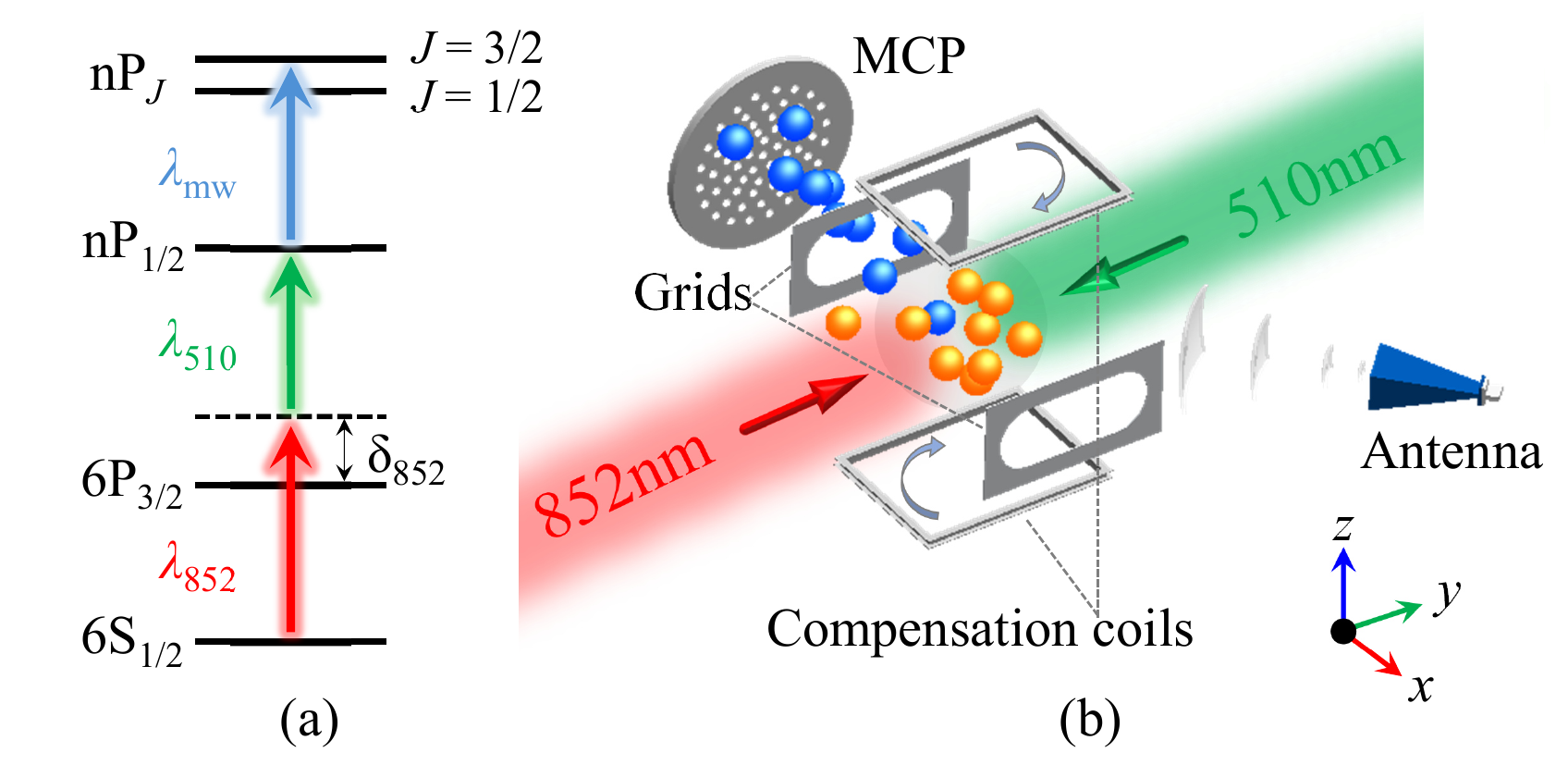}
	\caption{(a) Level diagram and excitation scheme. The $n \mathrm{S}_{1/2}$ state is excited by a two-step excitation with  $\lambda=852$-nm and 510-nm laser beams. First-step laser is blue-detuned $\delta_{852}/2\pi$ = +330~MHz from an intermediate state. Microwaves field $\lambda_{\mathrm{mw}}$ drives Rydberg transitions of $n\mathrm{S}_{1/2}\rightarrow n\mathrm{P}_{1/2,3/2}$. (b) Schematic of the experimental setup. The $\lambda=852$-nm and 510-nm excitation lasers are set to counter-propagate through the MOT center. A microwave field is emitted with an antenna to the MOT chamber and couples the $n\mathrm{S}_{1/2}$ $\to$ $n\mathrm{P}_{J}$ transition. Rydberg atoms are detected by the state-selective field ionization and time-gated ion detection with a MCP detector. Three pairs of grids and compensation coils are placed to compensate stray electric and magnetic fields (here only one pair of grids in $x$-axis and coils in $z$-axis are shown). Gold and blue balls represent laser-excited $n\mathrm{S}$ and microwave-coupled $n\mathrm{P}$ atoms, respectively.}
	\label{Fig1}
\end{figure}

The schematic of the experimental setup and related energy level are shown in Fig.~\ref{Fig1}.  A two-step excitation scheme is employed for exciting $ n\mathrm{S}_{1/2} $ state. The first-step laser is blue-detuned $\delta_{852}/2\pi=+330$~MHz from the intermediate $6\mathrm{P}_{3/2}$ state by a double-pass acousto-optic modulator (AOM) for eliminating photon scattering and radiation pressure. Then a microwave field is applied to couple the Rydberg transition of $ n\mathrm{S}_{1/2}  \to n\mathrm{P}_{1/2,3/2} $, see Fig.~\ref{Fig1}(a), forming narrow linewidth microwave spectra. 

Experiments are performed in a standard magneto-optical trap (MOT) with the peak density $\sim 10^{10} ~\text{cm}^{-3} $ and temperature of $\sim $100~$\mu$K, the main setup is shown in Fig.~\ref{Fig1}(b).  After switching off the MOT beams and waiting for a delay time of 1~ms, we turn on the Rydberg excitation pulse (500 ns) for populating the $n\mathrm{S_{1/2}}$ state and then a microwave pulse (30~$\mu$s ) for coupling the transition $ n\mathrm{S_{1/2}} \to n\mathrm{P_{J}} $ ($n = 41 - 55$). Both 852-nm and 1020-nm lasers are external-cavity diode lasers from Toptica that are locked to the 15000 finesse cavity, resulting in the laser linewidth less than 50~kHz. The 1020-nm laser is amplified and frequency doubled with the Precilasers (YFL-SHG-509-1.5) generating 510~nm second-step laser. 
Two lasers have Gaussian beam $1/e^2$ waist of $\omega_{852}$ = 750~$\mu$m and $\omega_{510}$ =1000~$\mu$m, respectively. The large beam waists and small laser intensities yield a small excitation Rabi frequency, $ \Omega = \Omega_{852}\Omega_{510}/(2\delta_{852}) \approx 2\pi \times 20$~kHz. For our 500~ns excitation pulse, the Rydberg excitation probability is quite small and the Rydberg atom density is less than $3\times 10^6 ~\text{cm}^{-3}$, which corresponds to atomic distance larger than 40~$\mu$m. The interaction induced shift between Rydberg atoms at this density is negligible in this work. 
The microwave field is generated by an analog signal generator (Keysight E8257D, frequency range 100 kHz to 67~GHz), and emitted with an antenna (A-INFO LB-15-15-c-185F, frequency range 50 to 65~GHz), covering the $ n\mathrm{S_{1/2}} \to n\mathrm{P_J} $ transition for $n=41-44$, and the antenna (A-INFO LB-180400-KF, frequency range 18 to 40 GHz) for $n=47-55$. 

After switching off the microwave field, we apply a ramped ionization electric field with ramp time 3~$\mu$s to ionize Rydberg atoms. The resultant Rydberg ions are collected with a microchannel plate (MCP) and sampled with a boxcar (SRS-250) and recorded with a computer.

\section{Microwave spectroscopy}

\begin{figure}[htpb]
\centering\includegraphics[width=0.45\textwidth]{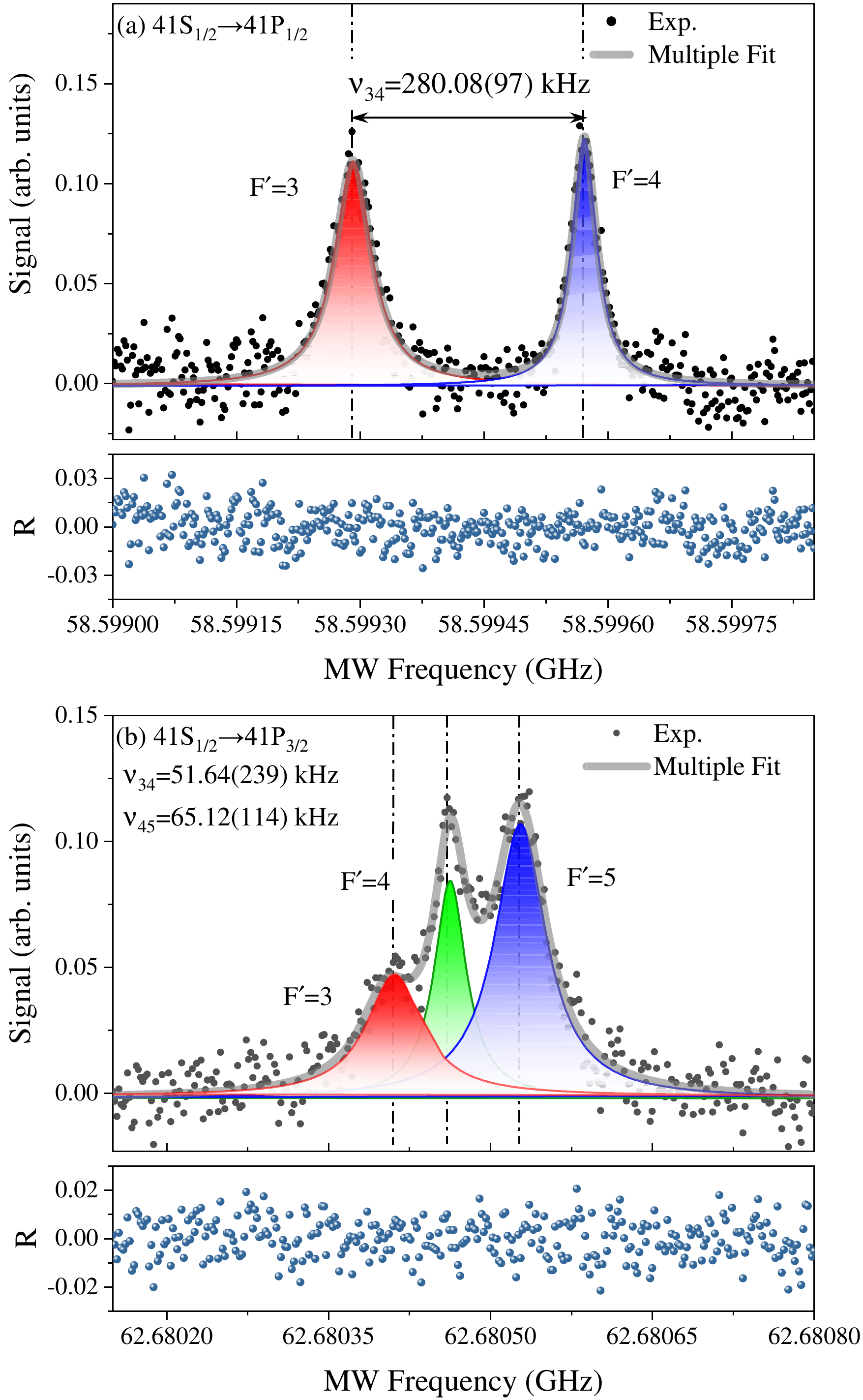}
\caption{ Measured microwave spectroscopy for a microwave field coupling the $41\mathrm{S}_{1/2} \rightarrow 41\mathrm{P}_{1/2}$  transition (a) and $41\mathrm{S}_{1/2} \rightarrow 41\mathrm{P}_{3/2}$ transition (b). Black dots show detected nP Rydberg signal, which are averaged over 25 repetitions of the experiment. Solid lines indicate multipeak Lorentz fittings, yielding the center frequency of hyperfine transition, marked with vertical dot-dashed lines, and further the hyperfine splitting of 41$\mathrm{P_J}$. The bottom plots display the residuals, R, of the Lorentzian fit and spectrum. Measured hyperfine splittings are $\nu_{34,\mathrm{P_{1/2}}} = 280.08(97)$~kHz for 41$\mathrm{P}_{1/2}$,  and $\nu_{34,\mathrm{P_{3/2}}}$ = 51.64(239)~kHz and $\nu_{45,\mathrm{P_{3/2}}}$= 65.12(114)~kHz for 41$\mathrm{P}_{3/2}$ state. Measured linewiths is 36.3~kHz  for the hyperfine transition line of $ n\mathrm{P}_{1/2},\mathrm{F}'=4 $, which approaches Fourier limit of 33~kHz. }
\label{Fig2}
\end{figure}

In the experiment, we lock the frequencies of the
two excitation lasers to resonantly prepare Rydberg
atoms in $n\mathrm{S}_{1/2}$ ($n = 41-55$) states.
The microwave frequency is scanned across the $n\mathrm{S}_{1/2}\rightarrow n\mathrm{P}_{1/2, 3/2}$ transitions. Fig.~\ref{Fig2} presents the measured microwave spectra of $41\mathrm{S}_{1/2} \to 41\mathrm{P}_{1/2}$ (a), and $41\mathrm{P}_{3/2}$(b) transitions.  The spectrum in Fig.~\ref{Fig2}(a) clearly shows two peaks, corresponding to the hyperfine transitions of $41\mathrm{S}_{1/2} (\mathrm{F=4}) \to 41\mathrm{P}_{1/2} (\mathrm{F'=3})$ and $(\mathrm{F'=4})$.  
The solid lines display Lorentz fits, yielding the center frequencies of 58.59929120(77)~GHz and 58.59957127(58)~GHz for $41\mathrm{S}_{1/2}(\mathrm{F=4})\rightarrow 41\mathrm{P}_{1/2},(\mathrm{F'=3})$ and $(\mathrm{F'=4})$, respectively. The statistical uncertainties, $\Delta \nu_{stat}$ in brackets, are less than 1~kHz. The extracted linewidths from Lorentz fits are 54.7~kHz and 36.3~kHz for $\mathrm{F'=3}$ and 4 lines, 
respectively, which approaches to the Fourier limit of 33.3~kHz for 30 $\mu s $-duration microwave pulse. From Fig.~\ref{Fig2}(a), we determine the hyperfine splitting $\nu_{34,\mathrm{P}_{1/2}}$ = 280.08(97)~kHz for $41\mathrm{P_{1/2}}$ state. Using a similar procedure, we obtain the microwave spectrum of the  $41\mathrm{S}_{1/2}\rightarrow 41\mathrm{P}_{3/2}$ transition, see Fig.~\ref{Fig2}(b). It is shown three peaks, corresponding to the hyperfine transitions of $41\mathrm{S}_{1/2}(\mathrm{F}=4)\rightarrow 41\mathrm{P}_{3/2}(\mathrm{F'=3,4,5})$.
From the multipeak Lorentz fitting shown with solid lines, we extract the center frequencies of hyperfine transitions marked with the vertical dot-dashed lines, and further the hyperfine splittings $\nu_{34,\mathrm{P_{3/2}}}= 51.64(239)$~kHz and $\nu_{45,\mathrm{P_{3/2}}} = 65.12(114)$~kHz. 
Due to the hyperfine structure being partially unresolved, the statistical uncertainty in the fitted line center is larger for $\mathrm{P_{3/2}}$ state than $\mathrm{P_{1/2}}$ state, and further the hyperfine splitting. In the bottom panels of Fig.~\ref{Fig2}, we plot the residuals, R, the difference between the data and the fitting.

\section{Systematic effects}
\label{sec:syst}

\begin{figure}[htpb]
	\centering\includegraphics[width=0.46\textwidth]{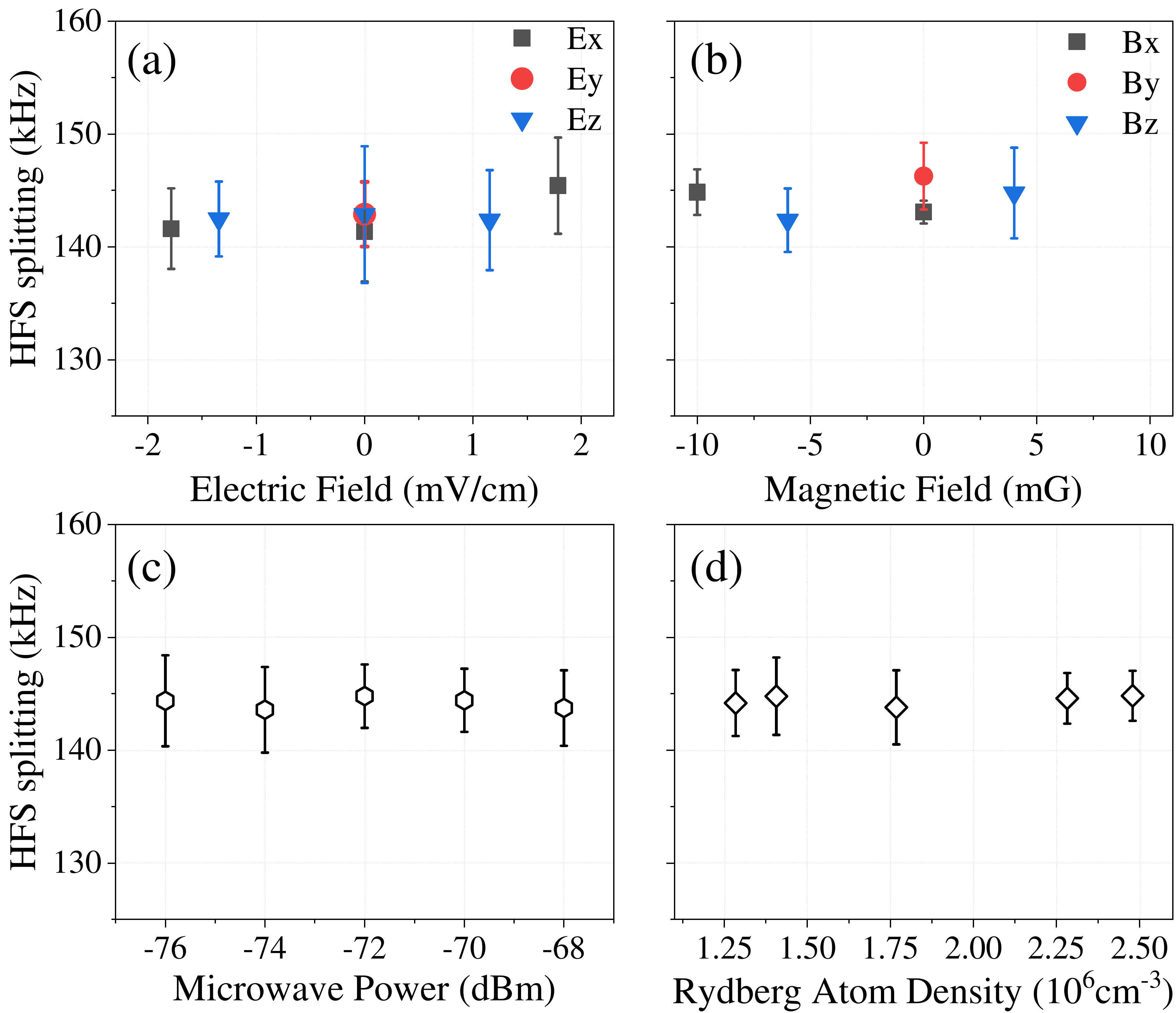}
	\caption{Measurements of the hyperfine splittings $\nu_{34,\mathrm{P_{1/2}}}$  of $51\mathrm{P_{1/2}}$ state for given applied weak electric field (a) and magnetic field (b) in all three directions. (c-d) measured $\nu_{34,\mathrm{P_{1/2}}}$ as a function of the microwave synthesizer output power (c) and Rydberg-atom density (d). The SEM of all measured $\nu_{34,\mathrm{P_{1/2}}}$ is used as the systematic uncertainty analysis. 
}
	\label{Fig3}
\end{figure}
 
 In this section, we take the spectrum of $ 51\mathrm{P}_{1/2} $ state as an example to analyze our  systematic uncertainties, including those arising from the microwave generator,
electric, and magnetic fields, as well as Rydberg-atom interactions.
The detailed uncertainties analysis is similar to our previous work~\cite{bai2023quantum}. From the measured microwave spectrum of Fig.~\ref{Fig2}, the statistical uncertainty is less than 1~kHz for $n\mathrm{P_{1/2}}$ and 2~kHz for $n\mathrm{P_{3/2}}$. Below we focus on the systematic shift and uncertainty. 

Firstly, we discuss the uncertainty of the signal generator frequency. To obtain accurate transition-frequency readings, we use an external atomic clock (SRS FS725) as a reference to lock the crystal oscillator of the microwave generator. The clock's relative uncertainty is $\pm~5\times10^{-11}$, leading the frequency deviation less than 10~Hz~\cite{moore2020measurement}. Therefore systematic shifts due to signal-generator frequency uncertainty are negligible. 

Secondly, we discuss the systematic uncertainty from stray electric and magnetic fields. During the experiments, we carefully zeroed the electric and magnetic field in the excitation area employing the Stark and Zeeman effect. After compensating, the stray electric and magnetic fields are less than 2~mV/cm and 5~mG, respectively~\cite{bai2023quantum,fan2024microwave}. These stray fields are sufficient to cause the line shift and broadening of the microwave spectrum. The symmetry of our observed spectral lines indicates that
background electric- and magnetic-field inhomogeneities can be negligible. In order to characterise the possible systematic uncertainties from any leakage within this range, we do the measurements of the hyperfine splitting $\nu_{34}$ for  $51\mathrm{S}_{1/2}\to 51\mathrm{P}_{1/2}$ transition by applying the weak electric and magnetic field in $x-$, $y-$ and $z-$ directions, that offsets the compensation electric field within $\pm$2~mV/cm and magnetic field within $\pm$10~mG, as shown in Fig.~\ref{Fig3}(a) and (b). We use the standard error of the mean (SEM) of the data points to estimate the systematic shift due to the stray electric and magnetic field in this work. A similar analysis was done in measurements of the $n\mathrm{P_{1/2}}$ hyperfine structure of $^{85}$Rb Rydberg states~\cite{cardman2022hyperfine}. For $n=51$ state in Fig.~\ref{Fig3}(a) and (b), the SEM analysis yields $\delta\nu_{34} =0.50$~kHz due to electric field and $\delta\nu_{34} =0.67$~kHz due to magnetic field. 

Thirdly, we discuss the systematic shift due to the AC Stark effect. The microwave intensity at the MOT center is
varied by changing the synthesizer output power. 
In Fig.~\ref{Fig3}(c), we display the measured hyperfine splitting   of the $51\mathrm{S}_{1/2}(\mathrm{F=4}) \rightarrow 51\mathrm{P}_{1/2},(\mathrm{F'=3,4})$ as a function of the microwave output power 
to evaluate systematic shifts due to the AC Stark effect. 
For the measured HFS frequency interval $\nu_{34}$ and the microwave power less than $-68$~dBm,
the measured transition frequency has no observable AC shift and the statistical variation of the HFS interval over this microwave power range is less than $\approx 0.23$~kHz. Therefore, we take multiple measurements within this power range and average the results to improve statistics.  

Finally, we consider the shift due to the interaction between Rydberg atoms. Rydberg energy level can be shifted due to the dipole-dipole and van-der-Waals interaction between Rydberg atoms~\cite{Gallagher1994}, which respective scale as $C_3/R^{3}$ and $C_6/R^{6}$ with $R$ the interatomic distance and $C_3$ and $C_6$ dispersion coefficients. The calculated dispersion coefficients are $C_6 \approx~14$~GHz $\mu \text{m}^6$ for  51$\mathrm{S}_{1/2}$  and $C_6 \approx -2$ GHz $\mu \text{m}^6$ for 51$\mathrm{P}_{1/2}$ atomic pair state. For our case of the atomic separation $R\approx 40~\mu$m, the level shifts due to Van der Waals interaction are a few Hz and are negligible.

In addition, atom pairs in a mix of $51\mathrm{S}_{1/2}$ and $51\mathrm{P}_{1/2}$ states strongly interact via resonant dipolar interaction, which scales as $C_3/R^3$. For $51\mathrm{S}_{1/2}-51\mathrm{P}_{1/2}$ pair states, the calculated dispersion coefficient is $|C_3|\approx 0.74$~GHz~$\mu \text{m}^3$, and the magnitudes of the line shifts are about 10~kHz. Considering the dipolar interaction potentials are symmetric about the asymptotic energies~\cite{bai2023quantum},
the main effect of the dipolar interaction is a line broadening without causing significant shift. 
In Fig.~\ref{Fig3}(d), we present  measurements of the hyperfine splitting as a function of the atomic density by varying the 
510-nm laser power. It can be seen that for the estimated Rydberg-atomic density of less than $3\times 10^6~\text{cm}^{-3}$, the density-induced line shift is less than 1~kHz. 

\section{Results and discussions}
\label{sec:results}
\subsection{Hyperfine splitting of \texorpdfstring{$n\mathrm{P}_{1/2}$}{nP 1/2} states}

We have performed a series of microwave-spectroscopy measurements like Fig.~\ref{Fig2}(a) for $n=41-55$. From these microwave spectra, we obtain the hyperfine splitting, $\nu_{34,\mathrm{P_{1/2}}}$,  and further the reduced hyperfine structure constant $\mathrm{A_{HFS,P_{1/2}}}$ using Eq.~(\ref{eq:A}), listed in the table~\ref{table I}. The quantum defects $\delta_0(\mathrm{P}_{1/2})=3.591556(30)$ and $\delta_2(\mathrm{P}_{1/2})=0.3714(40)$ are taken from Ref.~\cite{goy1982millimeterwave}. 
It is seen that the measured hyperfine splittings  $\nu_{34,\mathrm{P_{1/2}}}$ exhibit a significant decrease with the principal quantum number $n$, which is in agreement with the $n^{-3}$ scaling law. The statistical uncertainties $\delta\nu_{34,\mathrm{P_{1/2}}} \lesssim 1$~kHz for lower $n \lesssim 51$. Whereas the $\delta\nu_{34,\mathrm{P_{1/2}}}$ shows increasing for larger $n$, it is up to $\backsim $2~kHz for $n=55$. The larger statistical uncertainty for higher $n$ is mainly 
attributed to the decreased hyperfine splitting scaling as $n^{-3}$, our spectra can not resolve the hyperfine line well.

In the third column, we present the reduced hyperfine coupling constants and their uncertainties
using Eq.~3. 
Because the uncertainties of $\delta_0$ and $\delta_2$ lead to
shifts much smaller than our measurement uncertainties, we
neglect them in our uncertainty analysis. Therefore, $\delta \mathrm{A_{HFS}/A_{HFS}} =\delta\nu_{34}/\nu_{34}$. In the last two lines, we list the averaged reduced hyperfine coupling constant $\Bar{\mathrm{A}}_\mathrm{{HFS, P_{1/2}}} = 3.760$~GHz and statistical uncertainty $\delta \mathrm{A_{HFS,P_{1/2}}} =0.011$~GHz for $\mathrm{P}_{1/2}$ state.

\begin{table}[htpb]
\caption{ Summary of measured hyperfine splitting $\nu _{34,\mathrm{P_{1/2}}}$ and corresponding reduced hyperfine structure constants $\mathrm{A_{HFS,\mathrm{P_{1/2}}}}$  for $n = 41-55$. The number in bracket displays the statistical uncertainty.} 
\label{table I}
\begin{center}
	\renewcommand{\arraystretch}{1.7}
	\begin{tabular*}{\hsize}{@{}@{\extracolsep{\fill}}cccc@{}}
		\hline\hline
		$n$ & \quad $\nu_{34,\mathrm{P_{1/2}}} $~(kHz) \quad & \quad $\mathrm{A_{HFS,P_{1/2}}}$~(GHz)\quad \\
		\hline
		41 & \quad 280.08 (097) & \quad  3.665 (13) \\
		42 & \quad 263.86 (106) & \quad 3.738 (15) \\
		43 & \quad 240.35 (097) & \quad  3.677 (15) \\
		44 & \quad 220.61 (108) & \quad 3.639 (18) \\
		47 & \quad 185.85 (099) & \quad  3.800 (20) \\
		48 & \quad 173.35 (097) & \quad  3.795 (21) \\
		49 & \quad 161.83 (107) & \quad 3.788 (25) \\
		51 & \quad 143.08 (100) & \quad 3.811 (27) \\
		53 & \quad 128.31 (188) & \quad 3.869 (57) \\
		55 & \quad 112.41 (220) & \quad 3.818 (75) \\
		%57 & \quad 99.02 (300) & \quad 3.771 (115) \\
 \multicolumn{2}{l}{$\Bar{\mathrm{A}}_{\mathrm{HFS},P_{1/2}}$  (GHz) } & \quad3.760\\ 
 \multicolumn{2}{l}{Statistical uncertainty (GHz)}  & \quad 0.011\\
		\hline\hline
	\end{tabular*}
 \end{center}
\end{table}

Considering the systematic uncertainty analysis in section IV, we list the systematic uncertainty of $\delta A_{\mathrm{HFS,P_{1/2}}}$ in the table~\ref{table II}, including electric and magnetic field, ac Stark effect and Rydberg interaction induced uncertainty. Adding the systematic effect, the overall uncertainty of our measurement is $\delta \mathrm{A_{HFS, P_{1/2}}}=0.026$~GHz.

\begin{table}[htpb]
\caption{Uncertainty budget for measurements of $\mathrm{A_{HFS,P_{1/2}}}$, including systematic and  statistical uncertainties.}
\label{table II}
	\renewcommand{\arraystretch}{1.7}
	\begin{tabular*}{\hsize}{@{}@{\extracolsep{\fill}}ccc@{}}
		\hline\hline
		Source & \quad  $\delta \mathrm{A_{HFS,P_{1/2}}}$~(GHz) \quad \\
		\hline
		Electric Field          & \quad 0.013 \\
		Magnetic Field          & \quad 0.018 \\
		AC Stark                & \quad 0.006 \\
		Dipole-Dipole interactions  & \quad 0.005 \\
		Statistical uncertainty     & \quad 0.011   \\
		\hline\hline
	\end{tabular*}
\end{table}

\subsection{Hyperfine splitting of \texorpdfstring{$n\mathrm{P}_{3/2}$}{nP 3/2} states}

We do similar measurements of microwave spectra of $n\mathrm{P}_{3/2}$ states for $n=41-44$. We can not distinguish the hyperfine structure of $\mathrm{P}_{3/2}$ state for $n \geq 45$. From measured microwave spectrum like in Fig.~\ref{Fig2}(b), we extract the hyperfine splitting $\nu_{34,\mathrm{P_{3/2}}}$ and $\nu_{45,\mathrm{P_{3/2}}}$, 
listed in the table~\ref{table III}, 
with the statistical uncertainty in brackets. As expected, the hypefine splitting shows decrease with $n$ for both $\nu_{34,\mathrm{P_{3/2}}}$ and $\nu_{45,\mathrm{P_{3/2}}}$. 
Using Eqs.~(4) and (\ref{eq:ryd3}), we can determine the reduced magnetic-dipole HFS coupling constants $\mathrm{A_{HFS,P_{3/2}}}$ and electric-quadrupole HFS coupling constant $\mathrm{B_{HFS,P_{3/2}}}$, as well as their uncertainties, 
{with the quantum defect $\delta_0(\mathrm{P_{3/2}})=3.559058(30)$ and $\delta_2(\mathrm{P_{3/2}})=0.374(4)$, taken from  Ref.~\cite{goy1982millimeterwave}}.
In table~\ref{table III}, we also list extracted reduced hyperfine coupling constants of P$_{3/2}$ states and related statistical uncertainty. 
It is found that the uncertainty of $\nu_{34,\mathrm{P_{3/2}}}$ is about 2 times larger than that of $\nu_{45,\mathrm{P_{3/2}}}$, this is attributed to the hyperfine splitting of  $\nu_{34,\mathrm{P_{3/2}}}$ less than $\nu_{45,\mathrm{P_{3/2}}}$. For the higher $n$ state, such as 44P$_{3/2}$ state, the hyperfine splitting $\nu_{34,\mathrm{P_{3/2}}}$ is closer to spectral linewidth, therefore the multipeak Lorentz fitting may yield larger deviation for the hyperfine transition frequency of $n\mathrm{P_{3/2}, F'=3}$, and further splitting of $\nu_{34,\mathrm{P_{3/2}}}$. 
The larger deviation of $\nu_{34,\mathrm{P_{3/2}}}$ leads to the larger shift of 
B$_{\mathrm{HFS,P_{3/2}}}$ value, see results of n=43 and 44 in table~\ref{table III}. 

In last two lines of table~\ref{table III}, we list the averaged $\Bar{\mathrm{A}}_{\mathrm{HFS,P_{3/2}}}$ and $\Bar{\mathrm{B}}_{\mathrm{HFS,P_{3/2}}}$ value of this measurements. 
Considering the systematic effect of table~\ref{table II}, the overall uncertainties for our measurement of P$_{3/2}$ state are $\delta \mathrm{A_{HFS,P_{3/2}}}=0.027$~GHz and $\delta\mathrm{B_{HFS,P_{3/2}}}=0.102$~GHz, respectively.

\begin{table}[htb]
	\caption{Measured hyperfine splittings of $\mathrm{P}_{3/2}$ states and extracted reduced HFS coupling constant $\mathrm{A_{HFS,P_{3/2}}}$ and $\mathrm{B_{HFS,P_{3/2}}}$ by using Eqs.~(4) and (\ref{eq:ryd3}). 
  The hyperfine splittings $\nu_{34,\mathrm{P_{3/2}}}$ and $\nu_{45,\mathrm{P_{3/2}}}$ are in kHz and reduced coupling constants $\mathrm{A_{HFS,P_{3/2}}}$ and $\mathrm{B_{HFS,P_{3/2}}}$ are in GHz. 
{Number in bracket displays the statistical uncertainty.}}
	\label{table III}
	\begin{center}
		\renewcommand{\arraystretch}{1.7}
		\vspace{2ex}
		\begin{tabular}{ccccc}
			\hline\hline
			~$n$  &\quad $\nu_{34,\mathrm{P_{3/2}}}$ &\quad $\nu_{45,\mathrm{P_{3/2}}}$&\quad $\mathrm{A_{HFS,P_{3/2}}}$ &\quad $\mathrm{B_{HFS,P_{3/2}}}$  \\
			\hline
			41 & \quad 51.64 (239) & \quad 65.12 (114)  & \quad  0.680(21)  & \quad  0.028(157)  \\
			42 & \quad 50.53 (149) & \quad 62.71 (099)  & \quad  0.716(15)  & \quad  -0.024(112)\\
			43 & \quad 49.38 (383) & \quad 59.13 (165)  & \quad  0.747(40)  & \quad  -0.149(290)\\
			44 & \quad 44.96 (220) & \quad 53.07 (147)  & \quad  0.730(25)  & \quad  -0.193(192)\\
            \multicolumn{2}{c}{$\Bar{\mathrm{A}}_{\mathrm{HFS},P_{3/2}}$  (GHz) } &  & \quad0.718(13)  & \\
            \multicolumn{2}{c}{$\Bar{\mathrm{B}}_{\mathrm{HFS},P_{3/2}}$  (GHz) } &  & & \quad-0.084(099)  \\
        \hline\hline
		\end{tabular}
		\vspace{-3ex}
	\end{center}
\end{table}

\subsection{Comparison and Discussions}

\begin{figure}[htbp]
\centering\includegraphics[width=0.40\textwidth]{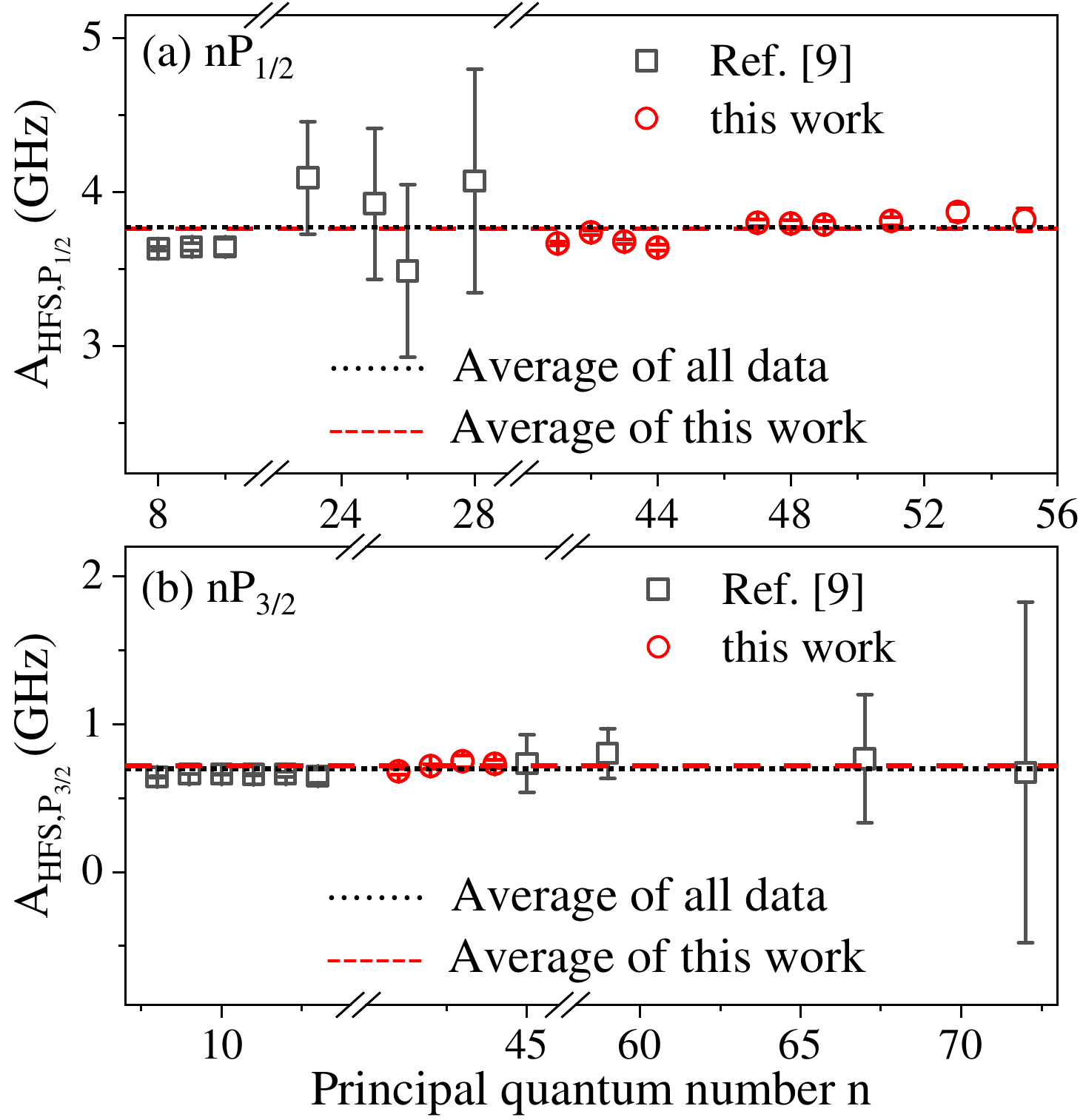}
	\caption{Comparisons of measured reduced hyperfine coupling constants, $\mathrm{A}_\mathrm{HFS,P_{1/2}}$   (a) and $\mathrm{A}_\mathrm{HFS,P_{3/2}}$ (b)  between this work and previous data in Ref.~\cite{allegrini2022survey} and reference therein. The horizontal black dot and red dashed lines display the averaged $\mathrm{A}_\mathrm{HFS,P_J}$ value for with and without literature data, respectively.  }
 \label{Fig4}
\end{figure}

For comparison with the literature data,  we display the reduced coupling constant 
$\mathrm{A_{HFS,P_{1/2}}}$ in Fig.~\ref{Fig4}(a) and  $\mathrm{A_{HFS,P_{3/2}}}$ in Fig.~\ref{Fig4}(b) with literature data taken from  Refs.~\cite{allegrini2022survey} and reference therein. Here we consider $n\mathrm{P_J}$ states with $n$ larger than 7. We can see that measurements of $\mathrm{A_{HFS, P_{J}}}$ were done mostly for lower principal quantum number $n\leqslant 13$ with an atomic beam or a vapor cell. 
In Ref.~\cite{goy1982millimeterwave},  $\mathrm{A_{HFS,P_{1/2}}}$ values of $n=23-28$ Rydberg states were measured with the cesium atomic beam and measured $\mathrm{A_{HFS,P_{1/2}}}$ has larger uncertainty.  In recent work~\cite{sassmannshausen2013highresolutionb},
$\mathrm{A_{HFS, P_{3/2}}}$ values for high lying Rydberg state with $n$ up to 72 were measured in the cesium MOT, where the $\mathrm{A_{HFS, P_{3/2}}}$ values were deduced from the line shape and line width, therefore with larger uncertainties. In this work,  we measure  $\mathrm{A_{HFS, P_{1/2}}}$ for $n=41-55$ range and $\mathrm{A_{HFS, P_{3/2}}}$ for $n=41-44$ range 
with the narrow linewidth microwave spectroscopy in MOT atoms. We analyse the hyperfine transition frequency of $n\mathrm{S}_{1/2}$(F=4) $\to$ $n$P$_\mathrm{J}(\mathrm{F'})$ and extract the reduced HFS constants with the smaller uncertainties, see tables~\ref{table I} and \ref{table III}.  The horizontal black dot and red dashed lines in Fig.~\ref{Fig4} display the averaged $\mathrm{A_{HFS, P_J}}$ value for with and without
literature data, respectively, which demonstrates our measurements agree with the literature values but with much less uncertainty. 

In addition, our measurements show reasonable agreement with the calculation data, where the HFS coupling constants of $\Bar{\mathrm{A}}_\mathrm{{HFS, P_{1/2}}}=3.666\pm0.007$ for $n = 8-17$, obtained by using the all-orders correlation potential method~\cite{grunefeld2019correlationa},  and $\Bar{\mathrm{A}}_\mathrm{{HFS, P_{1/2}}} =3.586 \pm 0.007$ and $\Bar{\mathrm{A}}_\mathrm{{HFS, P_{3/2}}} = 0.644 \pm 0.001$ for $n = 8-12$, calculated using the coupled-cluster with the single and double approximation 
 theory (CCSD)~\cite{tang2019initioa}. The error bar is the standard deviation. Note that the theoretical value of $\mathrm{A_{HFS, P_{J}}}$ here is the average of the reduced hyperfine coupling constant of $n$ state, which is extracted with the original data in the literature and $n^{*-3}$ scaling with the quantum defect in Ref.~\cite{goy1982millimeterwave}.  

{\color{red}
\begin{figure}[htbp]
\centering\includegraphics[width=0.40\textwidth]{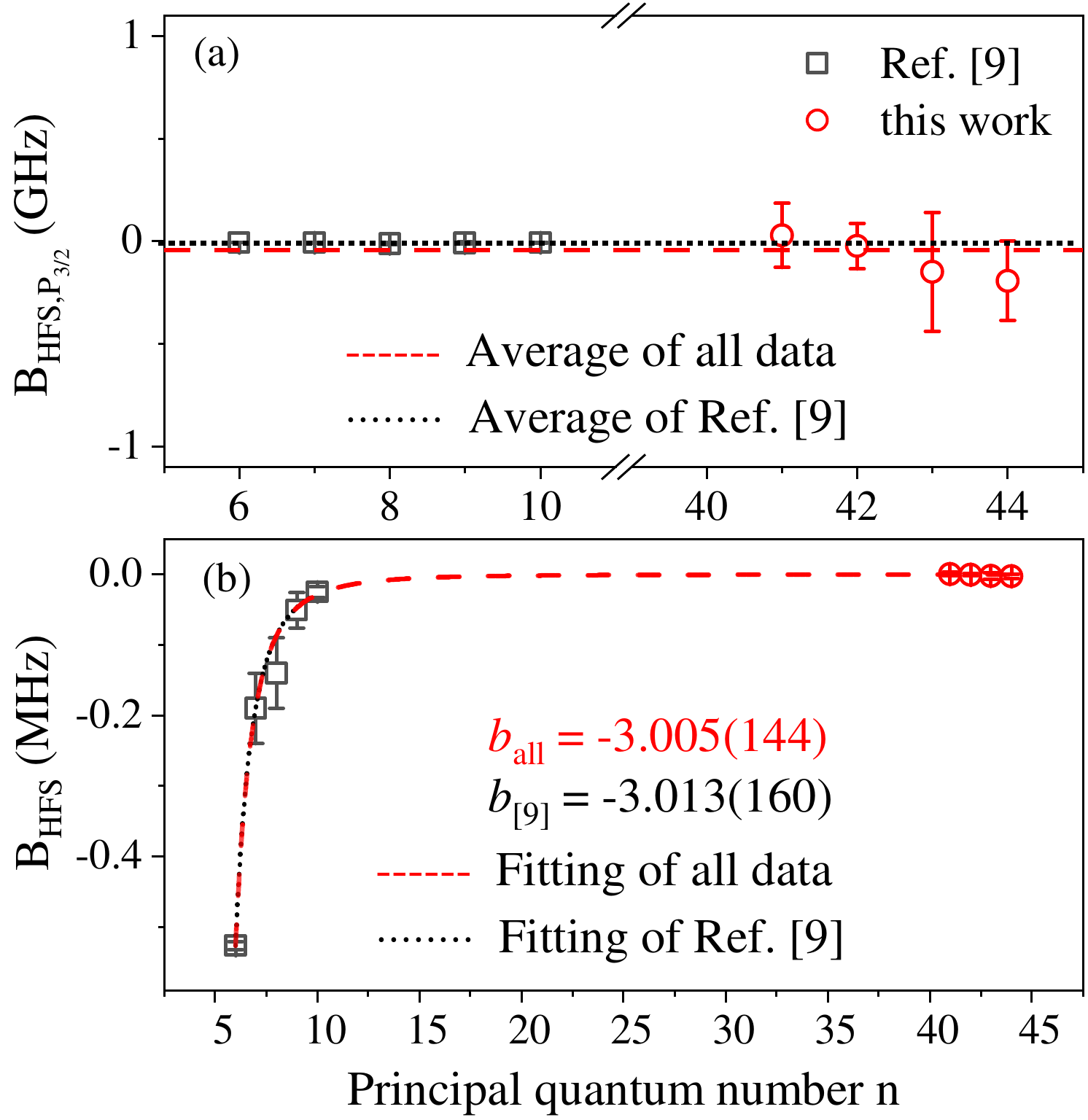}
	\caption{Comparison of measured reduced electric quadrupole constants, $\mathrm{B}_\mathrm{HFS,P_{3/2}}$ (a) and scaling test of HFS coupling constant $\mathrm{B}_\mathrm{HFS}$ (b)  between this work and previous data in Ref.~\cite{allegrini2022survey} and reference therein. The horizontal red dashed and black dot lines in (a) display the averaged $\mathrm{B}_\mathrm{HFS, P_J}$ value for all data and for literature data, respectively. Red dashed and black dotted lines in (b) display the allometric fittings of form $y=a\times n^{*b}$ for with and without our data, $a$ and $b$ are fitting parameters. }
 \label{Fig5}
\end{figure}}

 Finally, we discuss the electric-quadrupole HFS coupling constant. The  $\mathrm{B_{HFS,P_{3/2}}}$ value has been measured only for  $n\leq$~10 states with vapor cell in literature~\cite{allegrini2022survey}. In this work we measure the $\mathrm{B_{HFS,P_{3/2}}}$ value for $n = 41-44$ states with microwave spectra in the MOT.  
 To compare  $\mathrm{B_{HFS,P_{3/2}}}$ value of this work with literature data, in Fig.~\ref{Fig5}(a), we plot reduced $\mathrm{B_{HFS,P_{3/2}}}$ values versus $n$, where the previous data are extracted with the original data in Ref.~\cite{allegrini2022survey} and $n^{*-3}$ scaling with the quantum defect in Ref.~\cite{goy1982millimeterwave}. We can see that our results are consistent with the previous data but with a larger uncertainty specially for $n=43$ and 44, as the microwave spectrum for high $n$ state is partially unresolved, leading to larger deviation and uncertainty. In addition, 
 %to prove the feasibility of multipeak Lorentzians fitting to the $\mathrm{P}_{3/2}$ spectroscopy, 
 we also analyze the spectroscopy with the HF shfit of Eq.~(\ref{eq1}) and fit the spectrum of the transition $n\mathrm{S}_{1/2}(\mathrm{F}=4) \to n\mathrm{P}_{3/2}(\mathrm{F'})$ with the line shape model, see Appendix. 
 To explore the scaling law of $\mathrm{B_{HFS}}$, in Fig.~\ref{Fig5}(b), we present the
 $\mathrm{B_{HFS}}$ values as a function of the $n$, dashed and dotted lines show the allometric ($y=a\times n^{*b}$) fits for all data and literature data with fit parameters $b$ as indicated. Both fitting shows the $n^{*-3}$ scaling.

\section{Conclusion}

We have measured hyperfine structures and splittings of $n\mathrm{P_J} $ states with the narrow linewidth microwave spectroscopy. By analyzing the hyperfine splittings, we obtain reduced magnetic-dipole HFS coupling constant $ \mathrm{A_{HFS,P_J}} $ and electric-quadrupole HFS coupling constant $ \mathrm{B_{HFS,P_{3/2}}} $ for Rydberg $ n\mathrm{P_J} $ states. We have carefully analyzed systematic uncertainties due to the stray electric and magnetic field, the interaction between Rydberg atoms and the ac Stark effect.  Our measurements of magnetic-dipole constant $\mathrm{A_{HFS,P_J}}$ agree with previous data~\cite{allegrini2022survey,goy1982millimeterwave,sassmannshausen2013highresolutionb} 
and more precise.  
We also presented the reduced electric quadrupole coupling constant $ \mathrm{B_{HFS,P_{3/2}}} $ for cesium $n\mathrm{P}_{3/2}$ state.

Our precise measurement of intrinsic properties of Rydberg atoms, 
such as hyperfine coupling constant $\mathrm{A_{HFS}}$ and $\mathrm{B_{HFS}}$, is of significance for experimental investigation that relies on the availability of such data, as well as for testing of theoretical method.

\section{Acknowledgments}
We thank Prof. G. Raithel for his helpful discussion on experiments and calculations. This work is supported by the National Natural Science Foundation of China (Grant Nos. 12120101004, 62175136, 12241408 and U2341211); the Scientific Cooperation Exchanges Project of Shanxi province (Grant No. 202104041101015); 1331 project of Shanxi province. 

\appendix
\section*{Appendix: Line shape Model}
\renewcommand\theequation{A\arabic{equation}}
\setcounter{equation}{0}

\begin{figure}[htp!]
	\centering
	\includegraphics[width=0.95\linewidth]{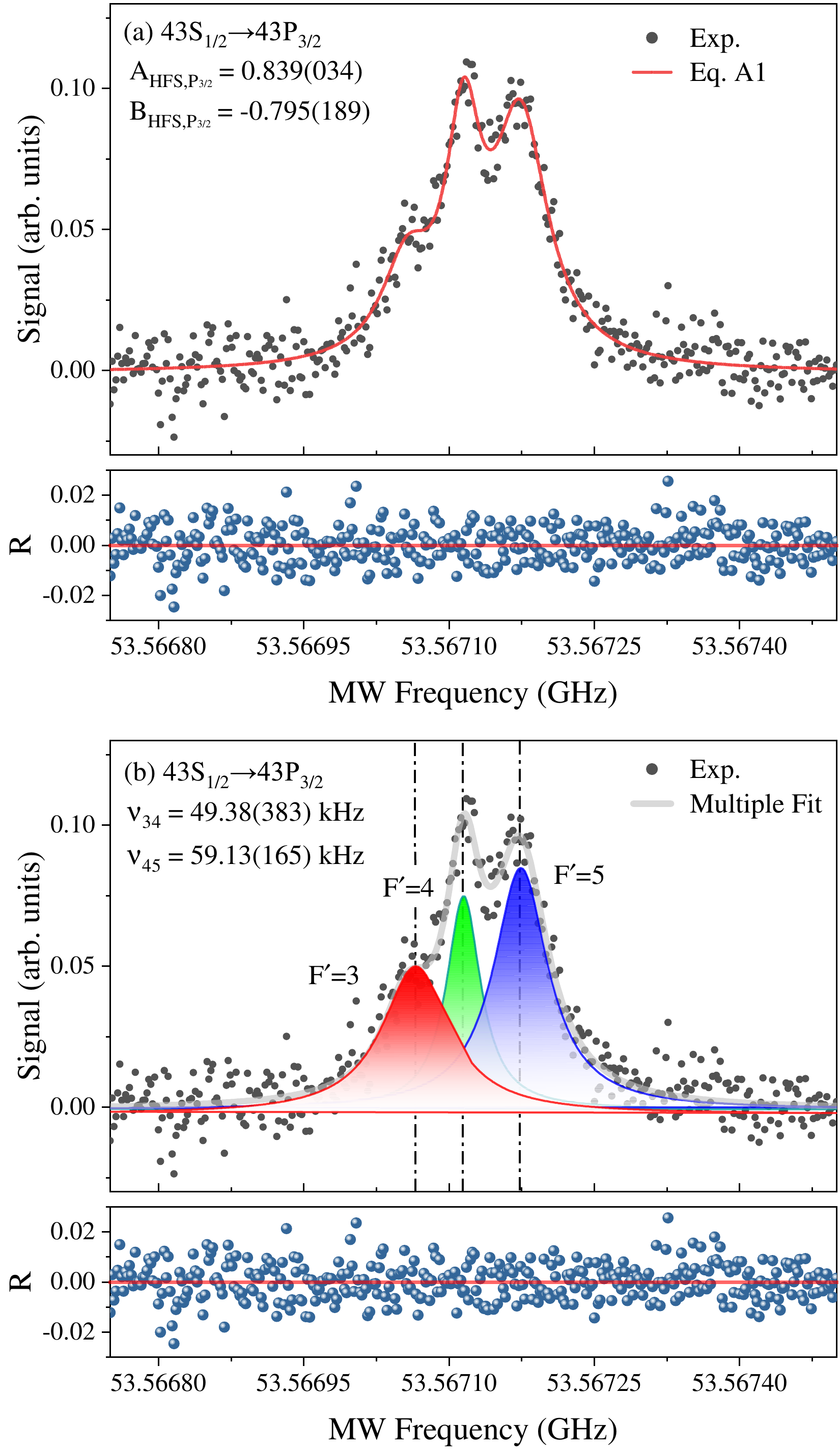}
	\caption{Hyperfine spectroscopy of 43$\mathrm{S_{1/2} (F=4)} \to$ 43$\mathrm{P_{3/2}(F’=3,4,5)} $ transition with multipeak fitting (a) and Eq.~\ref{eq:A1} fitting (b). The bottom panels display the residuals, R, of the fitting and spectrum.}
    \label{43P}
\end{figure} 
The shift of a hyperfine level from the center of gravity is determined with Eqs.~(\ref{eq1}) and ~(\ref{eq2}).
Considering the Lorentzian profile of spectrum for the transition S$_{1/2}(\mathrm{F}=4)$ $\to$ P$_{3/2}(\mathrm{F'})$, the fit function would be 

\begin{align}
\begin{split}
y=y_0&+\frac{2a_1}{\pi}\frac{\omega_1}{4(x-(x_{F=4\to cog}-\frac{15}{4}A-\frac{5}{28}B))^2+\omega_{1}^2}\\&+\frac{2a_2}{\pi}\frac{\omega_2}{4(x-(x_{F=4\to cog}+\frac{1}{4}A-\frac{13}{28}B))^2+\omega_{2}^2}\\&+\frac{2a_3}{\pi}\frac{\omega_3}{4(x-(x_{F=4\to cog}+\frac{21}{4}A+\frac{1}{4}B))^2+\omega_{3}^2},\\
\end{split}
\label{eq:A1}
\end{align}
where $a_i$ and $\omega_i$ $(i=1-3)$ represent the amplitude and width of the hyperfine peak, $x_{(F=4\to cog)}$ is the center of gravity of the transition. 
The frequency of every peak is determined by only the center of gravity and the hyperfine constants A and B. We fit the experimental spectrum with Eq.~\ref{eq:A1} and obtain the hyperfine constants A and B for $\mathrm{P_{3/2}}$ state. For example,  Fig.~\ref{43P}(a) displays the spectroscopy of $43\mathrm{S}_{1/2}\to 43\mathrm{P}_{3/2}$ transition and line shape model fitting, the  bottom panel is the calculated residual. For comparison, we also plot the multipeak Lorentz fitting result (in this work) in Fig.~\ref{43P}(b). Obtained reduced hyperfine constants A and B values using two methods are listed in Table~\ref{table A1}. It is seen that for the A value, the Eq.~\ref{eq:A1} fitting is essentially same and minor larger than the multipeak fitting. However, for the B value, the Eq.~\ref{eq:A1} fitting is larger than the multipeak fitting, which is probably because Eq.~\ref{eq:A1} has more fitting parameters and 
for fitting the spectrum, some parameters are given according to the spectroscopy, such as center of gravity $x_{(F=4\to cog)}$.  

\begin{table}[htb]
	\caption{Comparison of reduced values of $\mathrm{A_{HFS,P_{3/2}}}$ and $\mathrm{B_{HFS,P_{3/2}}}$ for $\mathrm{P}_{3/2}$ states, obtained by two fitting methods, with $\mathrm{A_{HFS,P_{3/2}}}$ and $\mathrm{B_{HFS,P_{3/2}}}$ in GHz. The number in the bracket displays the statistical uncertainty.}
	\label{table A1}
	\begin{center}
		\renewcommand{\arraystretch}{1.7}
		\vspace{2ex}
		\begin{tabular}{ccccc}
			\hline\hline
              ~$n$ & \multicolumn{2}{c}{Multipeak fitting} & \multicolumn{2}{c}{Formula(\ref{eq:A1}) fitting} \\
			 &\quad $\mathrm{A_{HFS,P_{3/2}}}$
            &\quad $\mathrm{B_{HFS,P_{3/2}}}$
            &\quad $\mathrm{A_{HFS,P_{3/2}}}$
            &\quad $\mathrm{B_{HFS,P_{3/2}}}$  \\
            \hline
			41 & \quad 0.680 (21) & \quad 0.028 (157)  & \quad  0.745(25)  & \quad  -0.429(160)  \\
			42 & \quad 0.716 (15) & \quad -0.024 (112)  & \quad  0.790(14)  & \quad  -0.544(094)\\
			43 & \quad 0.747 (40) & \quad -0.149 (290)  & \quad  0.839(34)  & \quad  -0.795(189)\\
			44 & \quad 0.730 (25) & \quad -0.193(192)  & \quad  0.766(34)  & \quad  -0.445(191)\\
            \hline\hline
		\end{tabular}
		\vspace{-3ex}
	\end{center}
\end{table}

%It is found that for the A value, the Eq.~\ref{eq:A1} fitting is very close to the multipeak fitting and minor larger than the multipeak fitting, however, for the B value, the Eq.~\ref{eq:A1} fitting is much larger than the multipeak fitting but with smaller uncertainty. This is probably because Eq.~\ref{eq:A1} has more fitting parameters and hard to fit the spectroscopy with all free parameters. For fitting the spectrum, some parameters are given according to the spectroscopy, such as the center of gravity $x_{F=4→cog}$ and $a_i$. Considering the unreasonable uncertainty of the Eq.~\ref{eq:A1} fitting, we use the multipeak fitting results in this work. }

\bibliography{reference}

\end{document}